\documentclass[a4paper,onecolumn,11pt,professionalfonts]{quantumarticle}
\pdfoutput=1
\usepackage[utf8]{inputenc}
\usepackage[english]{babel}
\usepackage[T1]{fontenc}
\usepackage{amsmath,amsthm,amssymb,hyperref,amssymb, amsfonts, latexsym}
\usepackage{breakurl}
\usepackage{amsthm}
\usepackage{mathtools}
\usepackage[dvipdfmx]{graphicx}
\usepackage[dvipsnames]{xcolor} 
\usepackage{bbm}
\usepackage{bm}
\usepackage{tikz}
\usepackage{lipsum}
\theoremstyle{plain}

\numberwithin{equation}{section}
\usepackage{indentfirst}
\usepackage{booktabs}
\usepackage{multirow}

\usepackage[backend=bibtex, style=numeric-comp,sorting=none]{biblatex} 
\bibliography{akanebibb}

\begin{document}
\title{CP conditions for GKSL-like master equations}

\author{Akane Watanabe}
\affiliation{Department of Physics, Waseda University, Tokyo 169-8555, Japan}
\orcid{0000-0003-4937-5826}
\email{aka-wa.eclips@ruri.waseda.jp}
\author{Takayuki Suzuki}
\affiliation{National Institute of Information and Communications Technology,
Nukui-Kitamachi 4-2-1, Koganei, Tokyo 184-8795, Japan}
\orcid{0000-0003-3400-976X}
\author{Makoto Unoki}
\affiliation{Department of Physics, Waseda University, Tokyo 169-8555, Japan}
\orcid{0000-0003-3052-3155}
\author{Hiromichi Nakazato}
\affiliation{Department of Physics, Waseda University, Tokyo 169-8555, Japan}
\orcid{0000-0002-5257-7309}

\begin{abstract}
The complete positivity (CP) of a quantum dynamical map (QDM) is, in general, difficult to show when its master equation (ME) does not conform to the Gorini-Kossakowski-Sudarshan-Lindblad (GKSL) form. 
The GKSL ME describes the Markovian dynamics, comprising a unitary component with time-independent Hermitian operators and a non-unitary component with time-independent Lindblad operators and positive time-independent damping rates.
Recently, the non-Markovian dynamics has received growing attention, and the various types of GKSL-like MEs with time-dependent operators are widely discussed; however, rigorous discussions on their CP conditions remain limited.
This paper presents conditions for QDMs to be CP, whose MEs take the GKSL-like form with arbitrary time dependence. 
One case considered is where its ME takes the time-local integro-differential GKSL-like form, which includes CP-divisible cases.
Another case considered is where the ME is time-non-local but can be approximated to be time-local in the weak-coupling regime.
As a special case of the time-non-local case, the same discussion holds for the time-convoluted GKSL-like form, which should be compared to previous studies.

\end{abstract}
\maketitle

\section{\label{sec:intro}Introduction}
The theory of open quantum systems is the essential framework for describing microscopic systems influenced by their surrounding environments \cite{Alicki, breuer2002theory, Rivas, Breuer2016}.
The dynamics of these systems is described by either a quantum dynamical map (QDM), which maps an arbitrary initial state to another state at an arbitrary time, or its differential equation, known as a master equation (ME).

This framework ensures the probabilistic interpretation of quantum mechanics, guaranteed by two fundamental characteristics of a QDM.
One characteristic is the complete positivity (CP), which ensures the positivity of probabilities when the system of interest is correlated with other systems. If a QDM has the Kraus representation \cite{Kraus, Choi1975}, the QDM is CP, but the CP is generally difficult to show in the form of MEs. 
The other characteristic is the trace-preserving (TP), which ensures that all probabilities always sum to unity. This property can be easily demonstrated by taking the trace of MEs.

The Gorini-Kossakowski-Sudarshan-Lindblad (GKSL) ME is one of the most significant MEs in the theory of open quantum systems \cite{GKS1976, L1976}.
This represents the most general form of a CPTP quantum dynamical semigroup, which is used to describe the Markovian dynamics.
The GKSL ME is time-local and consists of two parts: a unitary part with time-independent Hermitian operators and a non-unitary part with time-independent Lindblad operators and their coefficients, which are time-independent and positive damping rates.
The widely accepted derivation of the GKSL ME resorts to the Born-Markov approximation, which is valid for initially uncorrelated systems in the weak coupling regime \cite{breuer2002theory, Rivas}.
Since the standard derivation also involves the rotating wave approximation, finding alternative methods for deriving the GKSL ME remains an important problem \cite{CPforBrownianGKSL, GKSL2010, CPbyCGRed2019, Huang2023CP?}.
The GKSL ME has successfully described the Markovian dynamics, not only theoretically but also phenomenologically \cite{Gardiner2004}.

The non-Markovian effect has recently attracted researchers with increasing attention \cite{BreuerGKSLl2004, Kossakowski2009concrete, Kossakowski2009general, Chruscinski2014, Breuer2016, chruscinski2022, relanM2018, Breuer2004, Piilo2009, Piilo2010, ZT19's15, semiVacchini2013, Hall2014, VegaReviewNonMarkov2017}.
Due to the non-Markovianity, the QDM does not have the semigroup property and becomes time inhomogeneous \cite{chruscinski2022}.
Since the non-Markovian systems do not follow the GKSL ME, various alternative MEs have been proposed \cite{breuer2012GKSLlike, Anderson2014GKSLlike, OTA, ULE, IkedainSciPost2021, MozgunovCP2020, Alipour2020, trushechkin2021unifiedgksl}.
Some yield a time-local GKSL-like ME characterized by time-dependent Lindblad operators \cite{breuer2012GKSLlike, Anderson2014GKSLlike, OTA, ULE, IkedainSciPost2021, MozgunovCP2020}.

The CP property has been extensively studied for the Markovian dynamics, e.g., a recent research investigates the relation between the CP and relaxation rates \cite{ChruKim2021semigeneCP, ChruKim2021semirelaCP}.
In contrast, rigorous discussions of the CP property for the non-Markovian dynamics remain limited, particularly to the time-local cases \cite{Alicki1979, Schaller2008, MozgunovCP2020, Davidovic2020localGKSLl, linearityCPGKSL-like2022, Sugny2023}.
For the time-local cases, CP divisibility, i.e. whether a QDM can be divided into CP maps, is often studied \cite{wolf2008mathdivisible, GKSLlCGnonM2013, ChruscinskiCPdivisible2016399, Cabrera2019geoPCP, rgrlzedRed2023}.
For the time-non-local cases, positivity is discussed for the GKSL-like ME \cite{Wilkie2000, GKSL-like2008noCP}.
The general discussion of the CP condition is challenging, as exemplified by the introduction of an additional mathematical condition on the kernel \cite{Tarasov2021BPCP, Tarasov2021memory}.
In the time-convoluted case, the CP property has been proven perturbatively for the general form \cite{chruscinski2022} and for the GKSL-like form \cite{Kossakowski2009concrete, Kossakowski2009general, Chruscinski2014}.

This paper aims to rigorously describe the conditions for QDMs to be CP when their MEs take the GKSL-like form. Specifically, the integro-differential local and non-local GKSL-like MEs with arbitrary time dependence on a Hermitian operator of the unitary part and Lindblad operators of the non-unitary part are considered. The significance lies in evaluating the CP by obtaining explicit forms of QDMs as the solutions of the MEs.

\section{Setup and goals \label{sec:2setupgoals}}
\subsection{Quantum dynamical maps (QDMs) and complete positivity (CP) \label{sec:2-1}}
For a finite $d$-dimensional open quantum system, a quantum state is described by a $d\times d$ positive semidefinite and unit-trace density matrix.
The dynamics is described by a QDM, which is a CPTP superoperator $\Lambda_t$ mapping an arbitrary initial state $\rho$ to a state at a later time $t$: $\rho(t)=\Lambda_t\rho, \ \Lambda_0={\mathbbm 1}$, where ${\mathbbm 1}$ denotes the $d$-dimensional identity superoperator.

A QDM $\Lambda_t$ is TP iff a state at a later time $\rho(t)=\Lambda_t\rho$ is also unit-trace, or its ME is traceless.
A QDM $\Lambda_t$ is CP iff a superoperator $\Lambda_t\otimes{\mathbbm 1}$ is a positive semidefinite superoperator.
Here, the superoperator $\Lambda_t\otimes{\mathbbm 1}$ is positive semidefinite iff $(\Lambda_t\otimes{\mathbbm 1})|\Phi\rangle\langle \Phi|$ is positive semidefinite for an arbitrary positive state $|\Phi\rangle\langle \Phi|$ of a $2d$-dimensional system. 
Also, $\Lambda_t$ is CP iff $\Lambda_t$ has a Kraus representation: $\Lambda_t \rho= \sum_{\alpha} K_t^{\alpha}\rho K_t^{\alpha \dagger}, \ \sum_{\alpha}K_t^{\alpha \dagger}K_t^{\alpha}={\mathbbm 1}$.
In this paper, the CP of a map $\Lambda_t$ is evaluated by the positive semidefiniteness of a measure ${\cal M}(\Lambda_t):=\langle \Psi|\big(  \Lambda_t\otimes \mathbbm{1}_d\big)(|\Phi\rangle\langle \Phi|)|\Psi\rangle$, where $|\Psi\rangle$ is an arbitrary $2d$-dimensitonal vector.

\subsection{GKSL-like master equations \label{sec:2-2}}

Two types of integro-differential equations are discussed in this paper. The first is the time-local case for $\Lambda^{\rm l}_t$ and the second is the time-non-local case for $\Lambda^{{\rm nl}}_{t}$:
\begin{equation}
    \frac{d}{dt}\Lambda^{{\rm l}}_t\rho\equiv\int^t_0 dt' {\cal L}_{t,t'}^{\rm GKSL}  (\Lambda^{{\rm l}}_{t}\rho), \ \ \ \frac{d}{dt}\Lambda^{{\rm n}{\rm l}}_t\rho\equiv\int^t_0 dt' {\cal L}_{t,t'}^{\rm GKSL}  (\Lambda^{{\rm n}{\rm l}}_{t'}\rho).
\end{equation}
For both types, the GKSL-like kernel ${\cal L}^{\rm GKSL}_{t,t'}$ is considered to have an arbitrary time dependence on both $t$ and the integral variable $t'$. 
Note that the only difference between them appears in the time-dependence of its argument of the kernel.
More specifically, in this paper, we consider the GKSL-like kernel ${\cal L}^{\rm GKSL}_{t,t'}$ 
\begin{align}
    {\cal L}^{\rm GKSL}_{t,t'}\rho=   -i\Big[H'_{t,t'}, \rho \Big]+ \sum_i \Bigg(L^i_{t,t'}\rho L^{i \ \dagger}_{t,t'} - \frac{1}{2}\Big\{ L^{i \ \dagger}_{t,t'}L^i_{t,t'}, \rho  \Big\} \Bigg), \label{eq:me01}
\end{align}
where $H'_{t,t'}$ and $\{L^i_{t,t'} \}$ denote a Hermitian operator and Lindblad operators, respectively.
Notice that despite the time dependence of the Lindblad operators, the overall sign of the non-unitary part is positive. 
Previous studies showed that the discussion becomes simple when the GKSL-like kernel ${\cal L}^{\rm GKSL}_{t,t'}$ is divided into two superoperators ${\cal B}_{t,t'}, $ and $ {\cal Z}_{t,t'}$ with another operator $W_{t,t'}$\cite{Kossakowski2009concrete, Kossakowski2009general, Chruscinski2014}:
\begin{align}
    {\cal L}^{\rm GKSL}_{t,t'}\rho &=\Big({\cal B}_{t,t'} -{\cal Z}_{t,t'} \Big) \rho, \\
    {\cal B}_{t,t'}\rho&=\sum_i L^i_{t,t'} \ \rho \ L^{i \ \dagger}_{t,t'}, \\
    {\cal Z}_{t,t'}\rho&=  W_{t,t'} \ \rho+ \rho \   W_{t,t'}^{\dagger}, \  W_{t,t'}= iH'_{t,t'}+\frac{1}{2}\sum_i L^{i \ \dagger}_{t,t'}L^i_{t,t'}.
\end{align}
We may introduce two other maps $\Lambda^{\cal B}_t, \Lambda^{\cal Z}_t$ in association with ${\cal B}_{t,t'}$ and ${\cal Z}_{t,t'}$ for the local/non-local cases:
\begin{align}
    \frac{d}{dt}\Lambda^{{\rm l/nl},{\cal B}}_t\rho\equiv\int^t_0 dt' {\cal B}_{t,t'}  (\Lambda^{{\rm l/nl},{\cal B}}_{t/t'}\rho), \ \ \ \frac{d}{dt}\Lambda^{{\rm l/nl},{\cal Z}}_t\rho\equiv\int^t_0 dt'\Big( - {\cal Z}_{t,t'}  (\Lambda^{{\rm l/nl},{\cal Z}}_{t/t'}\rho) \Big).
\end{align}

To solve the time-non-local GKSL-like ME, one usually considers the weak-coupling regime.
The idea comes from the standard derivation of the GKSL ME, which starts from the Redfield equation \cite{Red1957}.
In the standard derivation of the Redfield equation, the coupling strength $g$ between the system of interest and the environment is assumed to be small, and only terms up to the second order $O(g^2)$ are kept. We note that the time-local and time-non-local Redfield equations describe the same dynamics up to the second order $O(g^2)$, which is shown in App.~\ref{app:localize}.

\subsection{Overview of the results \label{sec:2-3}}
This subsection provides a concise overview of the main achievements of this work. The time-local case is discussed in detail in Sec.~\ref{sec:local}, while the time-non-local case is examined in  Sec.~\ref{sec:nonlocal}.

For the time-local case, $\Lambda^{{\rm l},{\cal B}}_t$ is CP as shown in Subsec.~\ref{sec:localB}, $\Lambda^{{\rm l},{\cal Z}}_t$ is CP as shown in Subsec.~\ref{sec:localZ}, and therefore $\Lambda_t^{{\rm l}}$ is CP as shown in Subsec.~\ref{sec:localCP}. Specifically, if the time dependence of the kernel only appears through c-number functions, $\Lambda^{{\rm l}}_t$ is CP divisible, as shown in Subsec.~\ref{sec:CPdivis}. This result is a simple extension of the previous studies \cite{wolf2008mathdivisible, GKSLlCGnonM2013, ChruscinskiCPdivisible2016399,rgrlzedRed2023}.

For the time-non-local case, $\Lambda^{{\rm nl},{\cal B}}_t$ is CP as shown in Subsec.~\ref{sec:nonlocalB}.
Other than this, the following discussions only give the sufficient conditions for the maps to be CP. 
For $\Lambda^{{\rm nl},{\cal Z}}_t$, the sufficient condition is strict, as shown in Subsec.~\ref{sec:nonLZ}. With the perturbative approximation in the weak-coupling regime, $\Lambda^{{\rm nl},{\cal Z}}_t$ becomes CP as shown in Subsec.~\ref{sec:ZCP weak}. Therefore, in the weak-coupling regime, $\Lambda_t^{{\rm n}{\rm l}}$ is CP as shown in Subsec.~\ref{nonCPweak}.

Finally, the sufficient conditions for $\Lambda^{{\rm n}{\rm l}}_t$ are discussed under an additional assumption that $\Lambda^{{\rm nl},{\cal Z}}_t$ is CP. 
With the general time dependence of the kernel, the sufficient condition comes down to specific conditions on the Kraus representation of $\Lambda^{{\rm nl},{\cal Z}}_t$, as shown in Subsec.~\ref{sec:CPif}. 
This is also the case when the kernel is convoluted, as shown in Subsec.~\ref{sec:convovo}. This subsection also discusses the relations to the previous studies \cite{Kossakowski2009concrete, Kossakowski2009general, Chruscinski2014}.

\section{Time-local GKSL-like form \label{sec:local}}
\subsection{$\Lambda_t^{{{\rm l}},{\cal B}}$ is CP \label{sec:localB}}
Consider the first-order, linear integro-differential equation 
\begin{align}
    \frac{d}{dt}\Lambda^{{{\rm l}},{\cal B}}_t\rho=\int^t_0 dt' {\cal B}_{t,t'}  (\Lambda^{{{\rm l}},{\cal B}}_{t}\rho)=\int^t_0 dt'\Big( \sum_i L^i_{t,t'} (\Lambda^{{{\rm l}},{\cal B}}_{t}\rho) L^{i \ \dagger}_{t,t'}  \Big), \ \ \Lambda_0^{{{\rm l}},{\cal B}}={\mathbbm 1}.  \label{eq:NL2B1}
\end{align}
Integration of Eq.~(\ref{eq:NL2B1}) between $[0,t]$ explicitly shows the structure of $\Lambda^{{{\rm l}},{\cal B}}_t$
\begin{align}
    \Lambda_t^{{{\rm l}},{\cal B}}\rho= \rho &+ \int^t_0 dt_1 \int^{t_1}_0 dt_2 \sum_{i_1}L^{i_1}_{t_1,t_2} (\Lambda^{{{\rm l}},{\cal B}}_{t_1}  \rho)  L^{i_1 \ \dagger}_{t_1,t_2}. \label{eq:NL2B3} 
\end{align}
Note that the time locality in Eq.~(\ref{eq:NL2B1}) yields the $t_1$-dependence of $\Lambda^{{{\rm l}},{\cal B}}_{t_1}$ on the RHS of Eq.~(\ref{eq:NL2B3}).
Its formal solution is iteratively obtained
\begin{align}
    \Lambda_t^{{{\rm l}},{\cal B}}\rho= \rho + \sum_{n=1}^{\infty} \int^t_0 dt_1 \int^{t_1}_0 dt_2 \int^{t_1}_0 dt_3 \int^{t_3}_0 &dt_4\cdots \int^{t_{2n-3}}_0 dt_{2n-1}\int^{t_{2n-1}}_0 dt_{2n} \nonumber \\ 
    & \times \sum_{i_1, \cdots i_n} L^{i_1}_{t_1,t_2} \cdots L^{i_n}_{t_{2n-1},t_{2n}}    \ \rho \ L^{i_n \ \dagger}_{t_{2n-1},t_{2n}}  \cdots  L^{i_1 \ \dagger}_{t_1,t_2}.  \label{eq:NL2B2}        
\end{align}
The solution might appear complicated at first sight, but a simple example in App.~\ref{sec:example1} helps to see the essence.

The CP of $\Lambda_t^{{{\rm l}},{\cal B}}$ is shown by the positive semidefiniteness of ${\cal M}(\Lambda_t^{{{\rm l}},{\cal B}})$
\begin{align}
    {\cal M}(\Lambda_t^{{{\rm l}},{\cal B}})= |\langle \Psi|\Phi\rangle|^2+\sum_{n=1}^{\infty}  \int^t_0 dt_1 \cdots \int^{t_{2n-1}}_0 dt_{2n} \sum_{i_1, \cdots i_n}|\langle \Psi|\Big(L^{i_1}_{t_1,t_2} \cdots L^{i_n}_{t_{2n-1},t_{2n}}\Big) |\Phi\rangle|^2 \geq 0.
\end{align}

\subsection{$\Lambda_t^{{{\rm l}},{\cal Z}}$ is CP \label{sec:localZ}}
Consider the first-order, linear integro-differential equation 
\begin{align}
    \frac{d}{dt}\Lambda_t^{{{\rm l}},{\cal Z}}\rho =\int^t_0 dt' \Big(- {\cal Z}_{t,t'}  (\Lambda^{{{\rm l}},{\cal Z}}_{t}\rho)\Big)=\int^t_0 dt'\Big( -W_{t,t'} ( \Lambda_{t}^{{{\rm l}},{\cal Z}}  \rho ) -(\Lambda_{t}^{{{\rm l}},{\cal Z}}\rho ) W_{t,t'}^{\dagger}  \Big), \ \  \Lambda_0^{{{\rm l}},{\cal Z}}={\mathbbm 1}. \label{eq:NL2Z1}
\end{align}
Note that the equation is local in time. Therefore, the formal solution is obtained by using the chronological and anti-chronological exponential functions
\begin{align}
    \Lambda_t^{{{\rm l}},{\cal Z}}\rho= \overleftarrow{\cal T}e^{-\int^t_0ds\int^s_0ds'W_{s,s'}} \ \rho \ \overrightarrow{\cal T}e^{-\int^t_0ds\int^s_0ds' W_{s,s'}^{\dagger}} \equiv   V_{t} \rho V_{t}^{\dagger }.\label{eq:NL4Z2}
\end{align}
The explicit forms of $V_{t} $ and $ V_{t}^{\dagger}$ are the following
\begin{align}
    V_{t}& = 1+\sum_{n=1}^{\infty}\frac{(-1)^n}{n!}\int^t_0ds_1\int^{s_1}_0ds_2\cdots\int^t_0ds_{2n-1}\int^{s_{2n-1}}_0ds_{2n}\overleftarrow{\cal T}\big(W_{s_1,s_2} \cdots W_{s_{2n-1},s_{2n}}   \big), \\
    V_{t}^{\dagger}& =1+\sum_{n=1}^{\infty}\frac{(-1)^n}{n!}\int^t_0ds_1\int^{s_1}_0ds_2\cdots\int^t_0ds_{2n-1}\int^{s_{2n-1}}_0ds_{2n}\overrightarrow{\cal T}\big(W_{s_1,s_2}^{\dagger} \cdots W_{s_{2n-1},s_{2n}}^{\dagger}   \big).
\end{align}
Note that the chronological and anti-chronological orderings are applied to the odd number of the integral variables.

The CP $\Lambda_t^{{{\rm l}},{\cal Z}}$ is guaranteed by the positive semidefiniteness of the measure ${\cal M}(\Lambda_t^{{{\rm l}},{\cal Z}})$
\begin{align}
    {\cal M}(\Lambda_t^{{{\rm l}},{\cal Z}})= |\langle \Psi| V_{t} |\Phi\rangle|^2\geq 0.
\end{align}

\subsection{$\Lambda_t^{{\rm l}}$ is CP \label{sec:localCP}}
Consider the first-order, linear integro-differential, time-local GKSL-like master equation 
\begin{align}
    \frac{d}{dt}\Lambda^{{{\rm l}}}_t\rho=\int^t_0 dt' {\cal L}_{t,t'}^{\rm GKSL}  (\Lambda^{{{\rm l}}}_{t}\rho)=\int^t_0 dt'\Big( \sum_i L^i_{t,t'} (\Lambda^{{{\rm l}}}_{t}\rho) L^{i \ \dagger}_{t,t'}  -W_{t,t'} ( \Lambda_{t}^{{{\rm l}}}  \rho ) -(\Lambda_{t}^{{{\rm l}}}\rho ) W_{t,t'}^{\dagger}   \Big), \Lambda_0^{{{\rm l}}}={\mathbbm 1}.
    \label{eq:NL1}
\end{align}
To simplify Eq.~(\ref{eq:NL1}), take the time derivative of the map $\Lambda_t^{{\rm l}}$ under the similarity transformation $\Lambda_t^{{\rm l}}\equiv V_t \hat{\Lambda}_t^{{\rm l}} V_t^{\dagger}$
\begin{align}
    \frac{d}{dt}\Lambda_t^{{\rm l}}\rho=V_t\Big(\frac{d}{dt}\hat{\Lambda}_t^{{\rm l}}\rho\Big)V_t^{\dagger} + \int^t_0 dt' \Big( -W_{t,t'}V_t ( \hat{\Lambda}_{t}^{{{\rm l}}}  \rho )V_t^{\dagger} -V_t(\hat{\Lambda}_{t}^{{{\rm l}}}\rho )V_t^{\dagger} W_{t,t'}^{\dagger}   \Big). \label{eq:NL3}
\end{align}
Comparing Eq.~(\ref{eq:NL1}) and Eq.~(\ref{eq:NL3}), the first-order, linear integro-differential equation for $\hat{\Lambda}_t^{{\rm l}}$ is obtained
\begin{align}
    V_t\Big(\frac{d}{dt}\hat{\Lambda}_t^{{\rm l}}\rho\Big)V_t^{\dagger}=   \int^t_0 dt' \Big( \sum_i L^i_{t,t'} V_t(\hat{\Lambda}^{{{\rm l}}}_{t}\rho)V_t^{\dagger} L^{i \ \dagger}_{t,t'}\Big). \label{eq:NL4}
\end{align}
Define $\hat{L}^{i}_{s,s'}\equiv V_s^{-1}L^{i}_{s,s'}V_{s}$. See the explicit form of $V_t^{-1}$ in App.~\ref{sec:Vtinverse}.
Then, Eq.~(\ref{eq:NL4}) appears the same as Eq.~(\ref{eq:NL2B1})
\begin{align}
  \frac{d}{dt}\hat{\Lambda}_t^{{\rm l}}\rho=  \int^t_0 dt' \Big( \sum_i \hat{L}^i_{t,t'} (\hat{\Lambda}^{{{\rm l}}}_{t}\rho)\hat{L}^{i \ \dagger}_{t,t'}\Big). \label{eq:NL5}
\end{align}
Using the same arguments as in Subsec.~\ref{sec:localB}, Eq.~(\ref{eq:NL1}) is solved to give the formal solution
\begin{align}
    \Lambda_t^{{\rm l}} \rho = V_t\Bigg(\rho+ \sum_{n=1}^{\infty}  \int^t_0 dt_1 \cdots \int^{t_{2n-1}}_0 dt_{2n} \sum_{i_1, \cdots i_n} \hat{L}^{i_1}_{t_1,t_2} \cdots \hat{L}^{i_n}_{t_{2n-1},t_{2n}}    \ \rho \ \hat{L}^{i_n \ \dagger}_{t_{2n-1},t_{2n}}  \cdots  \hat{L}^{i_1 \ \dagger}_{t-1,t_2}       \Bigg)V_t^{\dagger}. \label{eq:TNLCP2}
\end{align}

The CP of $\Lambda_t^{{\rm l}}$ is guaranteed by the positive semidefiniteness of the measure ${\cal M}(\Lambda_t^{{\rm l}})$
\begin{align}
    {\cal M}(\Lambda_t^{{\rm l}})= |\langle \Psi|V_t|\Phi\rangle|^2 +\sum_{n=1}^{\infty}  \int^t_0 dt_1 \cdots \int^{t_{2n-1}}_0 dt_{2n} \sum_{i_1, \cdots i_n} |\langle \Psi| V_t \hat{L}^{i_1}_{t_1,t_2} \cdots \hat{L}^{i_n}_{t_{2n-1},t_{2n}} |\Phi\rangle|^2 
\geq 0 .\label{eq:grntCP}
\end{align}
Therefore, the QDM $\Lambda_t^{{\rm l}}$ for the time-local GKSL-like ME is CP.

\subsection{If the time dependence appears only through c-number functions, $\Lambda^{{{\rm l}}}_t$ is CP divisible \label{sec:CPdivis}}
Consider a special case of the time-local GKSL-like ME where the time dependence of the Lindblad operators is limited to c-number functions $h^i(t,t')\in{\mathbbm C}$, ${\rm i.e.}$, $L^i_{t,t'}=h^i(t,t')L^i$,
\begin{align}
    \frac{d}{dt}\Lambda^{{{\rm l}}}_t\rho=\int^t_0 dt'\Bigg(   -i\big[H'_{t,t'} , \Lambda^{{{\rm l}}}_t\rho \big] + \sum_i |h^i(t,t')|^2\Big( L^i (\Lambda^{{{\rm l}}}_{t}\rho) L^{i \dagger}  -\frac{1}{2} \big\{ L^{i \dagger}L^i, \Lambda_{t}^{{{\rm l}}}  \rho\big\} \Big)\Bigg), \ \Lambda_0^{{{\rm l}}}={\mathbbm 1}  \label{eq:NLCD1}.
\end{align}
The time-dependent damping rates $\int^t_0dt' |h^i(t,t')|^2$ are clearly positive semidefinite. This supports the well-known condition for the QDM to be CP-divisible \cite{wolf2008mathdivisible, GKSLlCGnonM2013, ChruscinskiCPdivisible2016399,rgrlzedRed2023}.

\section{Time-non-local GKSL-like form \label{sec:nonlocal}}

\subsection{$\Lambda_t^{{\rm n}{\rm l},{\cal B}}$ is CP \label{sec:nonlocalB}}
Consider the first-order, linear integro-differential equation 
\begin{align}
    \frac{d}{dt}\Lambda^{{\rm n}{\rm l},{\cal B}}_t\rho=\int^t_0 dt' {\cal B}_{t,t'}  (\Lambda^{{\rm n}{\rm l},{\cal B}}_{t'}\rho)=\int^t_0 dt'\Big( \sum_i L^i_{t,t'} (\Lambda^{{\rm n}{\rm l},{\cal B}}_{t'}\rho) L^{i \ \dagger}_{t,t'}  \Big)  , \ \ \Lambda_0^{{\rm n}{\rm l},{\cal B}}={\mathbbm 1}.  \label{eq:nonLB1}
\end{align}
Integration of Eq.~(\ref{eq:nonLB1}) between $[0,t]$ explicitly shows the structure of $\Lambda^{{\rm n}{\rm l},{\cal B}}_t$
\begin{align}
    \Lambda_t^{{\rm n}{\rm l},{\cal B}}\rho= \rho &+ \int^t_0 dt_1 \int^{t_1}_0 dt_2 \sum_{i_1}L^{i_1}_{t_1,t_2} (\Lambda^{{\rm n}{\rm l},{\cal B}}_{t_2}  \rho)  L^{i_1 \ \dagger}_{t_1,t_2}. 
     \label{eq:TNLB3} 
\end{align}
Note that the time non-locality is reflected on the $t_2$-dependence of $\Lambda^{{\rm n}{\rm l},{\cal B}}_{t_2}$ on the RHS of Eq.~(\ref{eq:TNLB3}).
Its formal solution is obtained
\begin{align}
    \Lambda_t^{{\rm n}{\rm l},{\cal B}}\rho= \rho + \sum_{n=1}^{\infty}  \int^t_0 dt_1 \int^{t_1}_0 dt_2 \int^{t_2}_0 &dt_3 \int^{t_3}_0 dt_4 \cdots \int^{t_{2n-2}}_0 dt_{2n-1}\int^{t_{2n-1}}_0 dt_{2n} \nonumber \\ 
    & \times \sum_{i_1, \cdots i_n} L^{i_1}_{t_1,t_2} \cdots L^{i_n}_{t_{2n-1},t_{2n}}    \ \rho \ L^{i_n \ \dagger}_{t_{2n-1},t_{2n}}  \cdots  L^{i_1 \ \dagger}_{t_1,t_2}.  \label{eq:nonLB2} 
\end{align}
Note that the intervals of integrations are naturally time-ordered and are different from those in Eq.~(\ref{eq:NL2B2}).

The CP of $\Lambda_t^{{\rm n}{\rm l},{\cal B}}$ is shown by the positive semidefiniteness of $ {\cal M}(\Lambda_t^{{\rm n}{\rm l},{\cal B}})$
\begin{align}
    {\cal M}(\Lambda_t^{{\rm n}{\rm l},{\cal B}})= |\langle \Psi|\Phi\rangle|^2+\sum_{n=1}^{\infty}  \int^t_0 dt_1 \cdots \int^{t_{2n-1}}_0 dt_{2n} \sum_{i_1, \cdots i_n}|\langle \Psi|\Big(L^{i_1}_{t_1,t_2} \cdots L^{i_n}_{t_{2n-1},t_{2n}}\Big) |\Phi\rangle|^2 \geq 0.
\end{align}

\subsection{ $\Lambda_t^{{\rm n}{\rm l},{\cal Z}}$ is CP only if $W_{t,t'}$ satisfies a strict condition \label{sec:nonLZ}}
Consider the first-order, linear integro-differential equation 
\begin{align}
    \frac{d}{dt}\Lambda_t^{{\rm n}{\rm l},{\cal Z}}\rho =\int^t_0 dt' \Big(- {\cal Z}_{t,t'}  (\Lambda^{{\rm n}{\rm l},{\cal Z}}_{t'}\rho)\Big)=\int^t_0 dt'\Big( -W_{t,t'} ( \Lambda_{t'}^{{\rm n}{\rm l},{\cal Z}}  \rho ) -(\Lambda_{t'}^{{\rm n}{\rm l},{\cal Z}}\rho ) W_{t,t'}^{\dagger}  \Big), \ \  \Lambda_0^{{\rm n}{\rm l},{\cal Z}}={\mathbbm 1} .\label{eq:nonLZ1}
\end{align}
In this subsection, the conditions for the solution $\Lambda_t^{{\rm n}{\rm l},{\cal Z}}$ to be CP are discussed. 
By integrating Eq.~(\ref{eq:nonLZ1}) between $[0, t]$, the structure of $\Lambda_t^{{\rm n}{\rm l},{\cal Z}}$ is explicitly shown
\begin{align}
    \Lambda_t^{{\rm n}{\rm l},{\cal Z}}\rho= \rho+\int^t_0 dt_1\int^{t_1}_0dt_2\Big(- {\cal Z}_{t_1,t_2}  (\Lambda^{{{\rm l}},{\cal Z}}_{t_2}\rho)  \Big).\label{eq:nonLZ2}
\end{align}
Iterating multiple times yields the formal solution as an infinite series expansion 
\begin{align}
    \Lambda_t^{{\rm n}{\rm l},{\cal Z}}\rho= \rho+ \int^t_0 dt_1\int^{t_1}_0dt_2 \Big( - {\cal Z}_{t_1,t_2}\rho \Big) + \int^t_0 dt_1\cdots\int^{t_3}_0dt_4\Big( {\cal Z}_{t_1,t_2} {\cal Z}_{t_3,t_4}\rho \Big)+\cdots.\label{eq:nonLZ3}
\end{align}

The condition for $\Lambda_t^{{\rm n}{\rm l},{\cal Z}}$ to be CP depends on the sign of ${\cal M}(\Lambda_t^{{\rm n}{\rm l},{\cal Z}})$:
\begin{align}
    {\cal M}(\Lambda_t^{{\rm n}{\rm l},{\cal Z}}) = |\langle\Psi|\Phi\rangle|^2&+\int^t_0 dt_1\int^{t_1}_0dt_2 \langle\Psi|\Big( - {\cal Z}_{t_1,t_2}|\Phi\rangle \langle\Phi| \Big)|\Psi\rangle \nonumber \\
    &+ \int^t_0 dt_1\cdots\int^{t_3}_0dt_4\langle\Psi|\Big( {\cal Z}_{t_1,t_2} {\cal Z}_{t_3,t_4}|\Phi\rangle \langle\Phi| \Big)|\Psi\rangle+\cdots. \label{eq:nonLZ4}
\end{align}
Showing the positive semidefiniteness of the measure ${\cal M}(\Lambda_t^{{\rm n}{\rm l},{\cal Z}})$ seems in general difficult. Since Eq.~(\ref{eq:nonLZ4}) is expressed as the power series expansion by the kernel $-{\cal Z}_{t,t'}$, the positive semidefiniteness of the second term of Eq.~(\ref{eq:nonLZ4}) is sufficient to guarantee $0\leq{\cal M}(\Lambda_t^{{\rm n}{\rm l},{\cal Z}})$.

It is shown, however, that the condition for the second term of Eq.~(\ref{eq:nonLZ4}) to be positive semidefinite imposes a strong condition on the operational structure of $W_{t,t'}$.
Express  $|\Psi\rangle=\sum_m\psi_m|m\rangle$, and an arbitrary initial state $\rho=\sum_n \rho_n|n\rangle\langle n|$ in terms of the complete orthonormal basis $\{|n\rangle\}$. The condition is then expressed as
\begin{align}
     0\leq \int^t_0 dt_1\int^{t_1}_0dt_2 \sum_{m,n,m'} \psi_m^* \rho_n\psi_{m'}\big( -\langle m| W_{t_1,t_2}|n\rangle \langle n|m'\rangle - \langle m|n\rangle \langle n|W_{t_1,t_2}^{\dagger}|m'\rangle\big).     \label{eq:nonLZ5}
\end{align}
This condition is indeed satisfied for arbitrary $\psi_m^* \rho_n\psi_{m'}$, by a trivial diagonal operator $W_{t_1, t_2}$ with some function $f(t_1,t_2)\in{\mathbbm C}$,
\begin{align}
    \langle m| W_{t_1,t_2}|n\rangle\propto -f(t_1,t_2)\delta_{n,m}, \ \  {\rm Re}\big[ f(t_1,t_2)  \big]<0. \label{eq:nonLZ52}
\end{align}
Actually, Eq.~(\ref{eq:nonLZ5}) is inevitably violated if the operator $W_{t_1,t_2}$ has a non-vanishing off-diagonal matrix element.
Consider a normalized vector $|\Psi\rangle=\psi_n|n\rangle+\psi_l|l\rangle$, and suppose $n\neq l$, $\langle l|W_{t_1,t_2}|n\rangle\neq 0$. 
Eq.~(\ref{eq:nonLZ5}) then becomes
\begin{align}
     0\leq \int^t_0 dt_1\int^{t_1}_0dt_2 \sum_{n}(-2\rho_n)\Big(|\psi_n|^2|\langle n| W_{t_1,t_2}|n\rangle|+|\psi_l\psi_n| |\langle l|W_{t_1,t_2}|n\rangle|\cos\big(\theta_n-\theta_l+\varphi_{l,n}\big)\Big),\label{eq:nonLZ6}
\end{align}
where relative phases $\varphi_{l,n}$ and $\theta_n-\theta_l$ are defined by $\langle l|W_{t_1,t_2}|n\rangle\equiv|\langle l|W_{t_1,t_2}|n\rangle|e^{i\varphi_{l,n}}$ and $\psi_l^*\psi_n\equiv|\psi_l||\psi_n|e^{i(\theta_n-\theta_l)}$.
Since $\rho_n$ of the initial state $\rho=\sum_n \rho_n|n\rangle\langle n|$ is positive semidefinite, Eq.~(\ref{eq:nonLZ6}) is expressed as below
    \begin{align}
      \sum_{n}\Bigg(\int^t_0 dt_1\int^{t_1}_0dt_2\frac{|\psi_n|}{\sqrt{1-|\psi_n|^2}}\frac{|\langle n| W_{t_1,t_2}|n\rangle|}{|\langle l|W_{t_1,t_2}|n\rangle|}+\cos\big(\theta_n-\theta_l+\varphi_{l,n}\big)\Bigg)\leq 0 .\label{eq:nonLZ7}
\end{align}
Although the condition by Eq.~(\ref{eq:nonLZ7}) has to be satisfied for arbitrary $\psi_n\in{\mathbbm C}$, this can always be violated by a certain choice of the relative phase $\theta_n-\theta_l$.
This concludes that no operator $W_{t_1,t_2}$ with non-vanishing off-diagonal elements is allowed for the condition Eq.~(\ref{eq:nonLZ5}) to hold.

In summary, for the first-order, linear integro-differential equation (\ref{eq:nonLZ1}), if the operator $W_{t_1, t_2}$ satisfies the strict condition Eq.~(\ref{eq:nonLZ52}), the map $\Lambda_t^{{\rm n}{\rm l},{\cal Z}}$ is CP.
The following subsections, therefore, impose additional constraints on the map $\Lambda_t^{{\rm n}{\rm l},{\cal Z}}$ to facilitate further analysis and discussion.

\subsection{$\Lambda^{{\rm n}{\rm l},{\cal Z}}_t$ is CP in the weak-coupling regime \label{sec:ZCP weak}}
Consider the first-order, linear integro-differential equation 
\begin{align}
    \frac{d}{dt}\Lambda_t^{{\rm n}{\rm l},{\cal Z}}\rho =\int^t_0 dt' \Big(- {\cal Z}_{t,t'}  (\Lambda^{{{\rm l}},{\cal Z}}_{t'}\rho)\Big)=\int^t_0 dt'\Big( -W_{t,t'} ( \Lambda_{t'}^{{{\rm l}},{\cal Z}}  \rho ) -(\Lambda_{t'}^{{{\rm l}},{\cal Z}}\rho ) W_{t,t'}^{\dagger}  \Big), \ \  \Lambda_0^{{\rm n}{\rm l},{\cal Z}}={\mathbbm 1} .\label{eq:nonLZ72}
\end{align}
In this subsection, the formal solution is perturbatively obtained. In the weak-coupling regime, the system-bath coupling strength $g$ is assumed to be sufficiently small and treated as a perturbation parameter. In the standard derivation of the GKSL-like ME, the Hermitian operator $H'_{t,t'}$ in the interaction picture and the Lindblad operators $\{L^i_{t,t'}\}$ are in proportion to $g^2$. The operator $W_{t,t'}$ is therefore proportional to $g^2$.

To further approximate Eq.~(\ref{eq:nonLZ72}), integrate Eq.~(\ref{eq:nonLZ72}) over $[0,t]$
\begin{align}
    \Lambda_t^{{\rm n}{\rm l},{\cal Z}}\rho =\rho + \int^t_0dt_1\int^{t_1}_0 dt_2\Big( -W_{t_1,t_2} ( \Lambda_{t_2}^{{\rm n}{\rm l},{\cal Z}}  \rho ) -(\Lambda_{t_2}^{{\rm n}{\rm l},{\cal Z}}\rho ) W_{t_1,t_2}^{\dagger}  \Big).\label{eq:nonLZ8}
\end{align}
Iterating Eq.~(\ref{eq:nonLZ8}) and neglecting the terms of order $g^3$ and higher, a simple differential equation is obtained
\begin{align}
    \frac{d}{dt}\Lambda_t^{{\rm n}{\rm l},{\cal Z}}\rho\approx \int^t_0dt' \Big( -W_{t,t'}  \rho  -\rho \ W_{t,t'}^{\dagger}  \Big).\label{eq:nonLZ9}
\end{align}
Since $W_{t_1,t_2}\propto g^2$, a further approximation: $g^2 \rho \approx g^2 \Lambda_t^{{\rm n}{\rm l},{\cal Z}}\rho$ is valid up to $g^2$. By this approximation, Eq.~(\ref{eq:nonLZ9}) appears as a familiar form 
\begin{align}
    \frac{d}{dt}\Lambda_t^{{\rm n}{\rm l},{\cal Z}}\rho\approx \Big(-\int^t_0dt' W_{t,t'}\Big) \Lambda_t^{{\rm n}{\rm l},{\cal Z}} \rho  +\Lambda_t^{{\rm n}{\rm l},{\cal Z}}\rho \Big(-\int^t_0dt' W_{t,t'}^{\dagger}  \Big).\label{eq:nonLZ10}
\end{align}
Eq.~(\ref{eq:nonLZ10}) coincides with Eq.~(\ref{eq:NL2Z1}). Thus the formal solution is the same as the time-local case
\begin{align}
    \Lambda_t^{{\rm n}{\rm l},{\cal Z}}\rho= \overleftarrow{\cal T}e^{-\int^t_0ds\int^s_0ds'W_{s,s'}} \ \rho \ \overrightarrow{\cal T}e^{-\int^t_0ds\int^s_0ds' W_{s,s'}^{\dagger}} \equiv   V_{t} \rho V_{t}^{\dagger }.\label{eq:NL4Z2}
\end{align}

The CP of $\Lambda_t^{{\rm n}{\rm l},{\cal Z}}$ is guaranteed by the positive semidefiniteness of the measure ${\cal M}(\Lambda_t^{{\rm n}{\rm l},{\cal Z}})$
\begin{align}
    {\cal M}(\Lambda_t^{{\rm n}{\rm l},{\cal Z}})= |\langle \Psi| V_{t} |\Phi\rangle|^2\geq 0.
\end{align}
In the weak-coupling regime, this type of map is CP.

\subsection{$\Lambda^{{\rm n}{\rm l}}_t$ is CP in the weak-coupling regime\label{nonCPweak}}
Consider the first-order, linear integro-differential, GKSL-like master equation 
\begin{align}
    \frac{d}{dt}\Lambda^{{\rm n}{\rm l}}_{t}\rho=\int^t_0 dt' {\cal L}_{t,t'}^{\rm GKSL}  (\Lambda^{{\rm n}{\rm l}}_{t'}\rho)&=\int^t_0 dt'\Big( \sum_i L^i_{t,t'} (\Lambda^{{\rm n}{\rm l}}_{t'}\rho) L^{i \ \dagger}_{t,t'}  -W_{t,t'} ( \Lambda_{t'}^{{\rm n}{\rm l}}  \rho ) -(\Lambda_{t'}^{{\rm n}{\rm l}}\rho ) W_{t,t'}^{\dagger}   \Big), \nonumber \\
    \Lambda_0^{{\rm n}{\rm l}}&={\mathbbm 1}.  \label{eq:nonlocalweak1}
\end{align}
In the weak-coupling regime, the discussion in Subsec.~\ref{sec:ZCP weak} holds, and Eq.~(\ref{eq:nonlocalweak1}) is approximated up to the second order of the system-bath coupling strength as
\begin{align}
    \frac{d}{dt}\Lambda^{{\rm n}{\rm l}}_{t}\rho\approx\int^t_0 dt'\Big( \sum_i L^i_{t,t'} (\Lambda^{{\rm n}{\rm l}}_{t'}\rho) L^{i \ \dagger}_{t,t'}  &-W_{t,t'} ( \Lambda_{t}^{{\rm n}{\rm l}}  \rho ) -(\Lambda_{t}^{{\rm n}{\rm l}}\rho ) W_{t,t'}^{\dagger}   \Big), \nonumber \\
    \Lambda_0^{{\rm n}{\rm l}}&={\mathbbm 1}.  \label{eq:nonlocalweak2}
\end{align}
In the same approach as in Subsec.~\ref{sec:localCP}, the contribution from the operator $W_{t,t'}$ can be incorporated by using the operators $V_t, V_t^{-1}$. 
Defining $\bar{L}^{i_j}_{s,s'}\equiv V_s^{-1}L^{i_j}_{s,s'}V_{s'}$, Eq.~(\ref{eq:nonlocalweak2}) takes the same form as Eq.~(\ref{eq:nonLB1})
\begin{align}
  \frac{d}{dt}\Lambda_t^{{\rm n}{\rm l}}\rho=  \int^t_0 dt' \Big( \sum_i \bar{L}^i_{t,t'} (\Lambda^{{{\rm l}}}_{t'}\rho)\bar{L}^{i \ \dagger}_{t,t'}\Big). \label{eq:NL5}
\end{align}
The formal solution of Eq.~(\ref{eq:nonlocalweak2}) is therefore 
\begin{align}
    \Lambda_t^{{\rm n}{\rm l}} \rho = V_t\Bigg(\rho+ \sum_{n=1}^{\infty}  \int^t_0 dt_1 \cdots \int^{t_{2n-1}}_0 dt_{2n} \sum_{i_1, \cdots i_n} \bar{L}^{i_1}_{t_1,t_2} \cdots \bar{L}^{i_n}_{t_{2n-1},t_{2n}}    \ \rho \ \bar{L}^{i_n \ \dagger}_{t_{2n-1},t_{2n}}  \cdots  \bar{L}^{i_1 \ \dagger}_{t_1,t_2}       \Bigg)V_t^{\dagger}. \label{eq:TNLCP2}
\end{align} 

The CP of $\Lambda_t^{{\rm n}{\rm l}}$ is guaranteed by the positive semidefiniteness of the measure ${\cal M}(\Lambda_t^{{\rm n}{\rm l}})$
\begin{align}
    {\cal M}(\Lambda_t^{{\rm n}{\rm l}})= |\langle \Psi|\Phi\rangle|^2 +\sum_{n=1}^{\infty}  \int^t_0 dt_1 \cdots \int^{t_{2n-1}}_0 dt_{2n} \sum_{i_1, \cdots i_n} |\langle \Psi| V_t \bar{L}^{i_1}_{t_1,t_2} \cdots \bar{L}^{i_n}_{t_{2n-1},t_{2n}} |\Phi\rangle|^2 \geq 0. \label{eq:grntCP}
\end{align}
Therefore, in the weak-coupling regime, $\Lambda_t^{{\rm n}{\rm l}}$ of the time-non-local GKSL-like ME is CP.

\subsection{$\Lambda_t^{{\rm n}{\rm l}}$ is CP if $\Lambda_t^{{\rm n}{\rm l},{\cal Z}}$ is CP and its Kraus representation has a specific property \label{sec:CPif}}

Consider the first-order, linear integro-differential, GKSL-like master equation 
\begin{align}
    \frac{d}{dt}\Lambda^{{\rm n}{\rm l}}_{t}\rho=\int^t_0 dt' {\cal L}_{t,t'}^{\rm GKSL}  (\Lambda^{{\rm n}{\rm l}}_{t'}\rho)&=\int^t_0 dt'\Big( \sum_i L^i_{t,t'} (\Lambda^{{\rm n}{\rm l}}_{t'}\rho) L^{i \ \dagger}_{t,t'}  -W_{t,t'} ( \Lambda_{t'}^{{\rm n}{\rm l}}  \rho ) -(\Lambda_{t'}^{{\rm n}{\rm l}}\rho ) W_{t,t'}^{\dagger}   \Big), \nonumber \\
    \Lambda_0^{{\rm n}{\rm l}}&={\mathbbm 1}.  \label{eq:nonlocalif1}
\end{align}
Assume $\Lambda_t^{{\rm n}{\rm l},{\cal Z}}$ which satisfies Eq.~(\ref{eq:nonLZ1}) is CP. Then, it has a Kraus representation $\Lambda_t^{{\rm n}{\rm l},{\cal Z}}\rho= \sum_j K_t^j \rho K_t^{j \dagger}$.
Further, assume that the Kraus operators satisfy specific conditions
\begin{align}
\big(K_t^{j}\big)^{-1}K_t^k={\mathbbm 1 }\delta_{j,k}, \ \  K_t^{j \dagger}\big(K_t^{k\dagger}\big)^{-1}={\mathbbm 1 }\delta_{j,k}. \label{eq:nonlocalif2}
\end{align}
This assumption is nothing but a technical requirement.

Consider another map $\hat{\Lambda}_t^{{\rm n}{\rm l}}$ which is defined by the similarity transformation of $\Lambda_t^{{\rm n}{\rm l}}$
\begin{align}
    \Lambda_t^{{\rm n}{\rm l}}\rho\equiv\sum_j K_t^j \Big( \hat{\Lambda}_t^{{\rm n}{\rm l}}\rho\Big) K_t^{j \dagger}. \label{eq:nonlocalif4}
\end{align}
Comparing the time derivative of Eq.~(\ref{eq:nonlocalif4}) and Eq.~(\ref{eq:nonlocalif1}), a similar form to Eq.~(\ref{eq:nonLB1}) is obtained for $\hat{\Lambda}_t^{{\rm n}{\rm l}}$
\begin{align}
   \sum_j K_t^j\Big(\frac{d}{dt}\hat{\Lambda}_t^{{\rm n}{\rm l}}\rho\Big)K_t^{j \dagger}=   \int^t_0 dt' \Big( \sum_i L^i_{t,t'}\big( \sum_k K_{t'}^k(\hat{\Lambda}^{{\rm n}{\rm l}}_{t'}\rho)K_{t'}^{k \dagger}\big) L^{i \ \dagger}_{t,t'}\Big). \label{eq:nonlocalif5}
\end{align}
Using the assumed properties of the Kraus operators and defining $\bar{L}^{i,j,k}_{s,s'}\equiv \big(K_s^j\big)^{-1}L^{i}_{s,s'}K_{s'}^k$, Eq.~(\ref{eq:nonlocalif5}) appears the same as Eq.~(\ref{eq:nonLB1})
\begin{align}
  \frac{d}{dt}\hat{\Lambda}_t^{{\rm n}{\rm l}}\rho=  \int^t_0 dt' \Big( \sum_{i}\sum_{j,k} \bar{L}^{i,j,k}_{t,t'} (\hat{\Lambda}^{{\rm n}{\rm l}}_{t'}\rho)\bar{L}^{i,j,k \dagger}_{t,t'}\Big). \label{eq:nonlocalif6}
\end{align}
According to Subsec.~\ref{sec:nonlocalB}, Eq.~(\ref{eq:nonlocalif1}) also has the formal solution
\begin{align}
    \Lambda_t^{{\rm n}{\rm l}} \rho =\sum_m K^m_t\Bigg(\rho+ \sum_{n=1}^{\infty}  \int^t_0 dt_1 &\cdots \int^{t_{2n-1}}_0 dt_{2n} \sum_{i_1, \cdots i_n}\sum_{j_1, \cdots j_n}\sum_{k_1, \cdots k_n}\nonumber \\
    &\times\bar{L}^{i_1,j_1,k_1}_{t_1,t_2} \cdots \bar{L}^{i_n,j_n,k_n}_{t_{2n-1},t_{2n}}    \ \rho 
     \bar{L}^{i_n,j_n,k_n \ \dagger}_{t_{2n-1},t_{2n}}  \cdots  \bar{L}^{i_1,j_1,k_1  \dagger}_{t_1,t_2}       \Bigg)K_t^{m \dagger}. \label{eq:TNLCP2}
\end{align}

The CP of $\Lambda_t^{{\rm n}{\rm l}}$ is guaranteed by the positive semidefiniteness of the measure ${\cal M}(\Lambda_t^{{\rm n}{\rm l}})$
\begin{align}
    {\cal M}(\Lambda_t^{{\rm n}{\rm l}})= \sum_m\Bigg( |\langle \Psi|K_t^m|\Phi\rangle|^2 +\sum_{n=1}^{\infty}  \int^t_0 dt_1 &\cdots \int^{t_{2n-1}}_0 dt_{2n} \sum_{i_1, \cdots i_n}\sum_{j_1, \cdots j_n}\sum_{k_1, \cdots k_n} \nonumber \\
    &\times |\langle \Psi| K_t^m \bar{L}^{i_1,j_1,k_1}_{t_1,t_2} \cdots \bar{L}^{i_n,j_n,k_n}_{t_{2n-1},t_{2n}}|\Phi\rangle|^2 \Bigg)\geq0. \label{eq:grntCP}
\end{align}
Therefore, the QDM $\Lambda_t^{{\rm n}{\rm l}}$ for the time-non-local GKSL-like ME is CP if its associated map $\Lambda_t^{{\rm n}{\rm l}, {\cal Z}}$ is CP and its Kraus operators satisfy the specific conditions.

\subsection{$\Lambda^{\rm conv}_t$ is CP if $\Lambda_t^{{\rm conv},{\cal Z}}$ is CP \label{sec:convovo}}
As a special case, consider the first-order, linear integro-differential GKSL-like master equation with convolution
\begin{align}
    \frac{d}{dt}\Lambda^{\rm conv}_{t}\rho=\int^t_0 dt' {\cal L}_{t-t'}^{\rm GKSL}  (\Lambda^{\rm conv}_{t'}\rho), \ \ \ 
    \Lambda_0^{\rm conv}&={\mathbbm 1}.  \label{eq:convo1}
\end{align}
Assume that the associated map $\Lambda^{{\rm conv},{\cal Z}}_{t}$ is CP and its Kraus representation has a specific property: $ \Lambda_t^{{\rm conv},{\cal Z}}\rho= \sum_j K_t^j \rho K_t^{j \dagger}, \big(K_t^{j}\big)^{-1}K_t^k={\mathbbm 1 }\delta_{j,k},\  K_t^{j \dagger}\big(K_t^{k\dagger}\big)^{-1}={\mathbbm 1 }\delta_{j,k}.$
Then, the formal solution of Eq.~(\ref{eq:convo1}) can be iteratively obtained and shown to be CP.

A simpler discussion is given in the previous studies \cite{Kossakowski2009concrete, Kossakowski2009general, Chruscinski2014}. In these works, the CP of $\Lambda^{\rm conv}_{t}$ is shown when the associated maps $\Lambda^{{\rm conv},{\cal B}}_{t} $ and $ \Lambda^{{\rm conv},{\cal Z}}_{t}$ are both CP. Since the proof relies on the Laplace transform, the explicit form of $\Lambda^{{\rm conv},{\cal Z}}_{t}$ is irrelevant, and therefore, an additional condition is not necessary.

\section{Conclusion \label{sec:conclusion}} 
This paper has examined the conditions for QDMs to be CP when their MEs take the GKSL-like form.
In the time-local case, a QDM is CP with arbitrary time dependence in the GKSL-like kernel of the integro-differential ME, which signifies that the GKSL-like form is both the necessary and sufficient condition for a QDM to be CP.
In the time-non-local case, an additional requirement is necessary for QDMs to be CP.
The condition has been proposed in the form of a specific Kraus representation of a partial map or the weak coupling regime.
The difference between the two cases lies in the locality, which arises from the Born approximation in the derivation in the weak coupling regime.

\section*{Acknowledgement}
T.S. was supported by Japan’s MEXT Quantum Leap Flagship Program Grant No. JPMXS0120319794.

\appendix
\section{Approximation in the derivation of the Redfield equation \label{app:localize}}
\renewcommand{\theequation}{A.\arabic{equation}}
\setcounter{equation}{0}
This section is aimed at showing that the time-local Redfield equation describes the same dynamics as its time-non-local counterpart in the weak-coupling regime. In the standard derivation, the microscopic Hamiltonian is given by $H=H_{\rm S}+H_{\rm B}+gSB$, where $H_{\rm S},\ H_{\rm B},\ g,\ S,$ and $ B$ denote the system free Hamiltonian, the bath free Hamiltonian, the coupling strength, the system part of the interaction, and the bath part of the interaction, respectively. In the weak-coupling regime, the coupling strength $g$ is small and only terms up to the second order have been kept. The initial state is assumed to be decoupled from the bath, and the system state is taken as an arbitrary state $\rho(0)$, while the bath state is taken as a thermal equilibrium $\rho_{\rm B}$. 

In the standard derivation of the Redfield equation, an integro-differential equation in the interaction picture is derived
\begin{align}
    \frac{d}{dt}\rho^{\rm I}(t)= g^2 \int^t_0d\tau \Big( 
    -S^{\rm I}(t)C(\tau)S^{\rm I}(-\tau)\rho^{\rm I}(t-\tau)
    +S^{\rm I}(t)\rho^{\rm I}(t-\tau)C^{*}(\tau)S^{\rm I}(-\tau) 
    \Big) +{\rm h.c.} , \label{eq:Red1}
\end{align}
where $C(\tau)={\rm Tr_B}\big[B^{\rm I}(t)B\rho_{\rm B} \big]$ denotes the correlation function.
The concrete expression of the first term of Eq.~(\ref{eq:Red1}) shows the $g$ dependence 
\begin{align}
    &-g^2\int^t_0d\tau S^{\rm I}(t)C(\tau)S^{\rm I}(-\tau)\rho^{\rm I}(t-\tau) \\
    &=-g^2\int^t_0d\tau S^{\rm I}(t)C(\tau)S^{\rm I}(-\tau)e^{iH_{\rm S}(t-\tau)}{\rm Tr_B}\big[e^{iH(t-\tau)}(\rho(0)\otimes\rho_{\rm B})e^{-iH(t-\tau)} \big]e^{-iH_{\rm S}(t-\tau)}.
\end{align}
Since only the terms up to the second order of $g$ are kept, any $g$-dependent parts in the partial trace can be neglected
\begin{align}
    &{\rm Tr_B}\big[e^{-iH(t-\tau)}(\rho(0)\otimes\rho_{\rm B})e^{iH(t-\tau)} \big] ={\rm Tr_B}\big[e^{iH\tau}\rho_{\rm tot}(t)e^{-iH\tau} \big]  \nonumber \\
    &= {\rm Tr_B}\Big[\big(1+i(H_{\rm S}+H_{\rm B}+gSB)\tau-\cdots\big)\rho_{\rm tot}(t)\big(1-i(H_{\rm S}+H_{\rm B}+gSB)\tau-\cdots\big)\Big] \nonumber \\
    &\approx {\rm Tr_B}\Big[e^{iH_{\rm S}\tau}\rho_{\rm tot}(t)e^{-iH_{\rm S}\tau}\Big] =e^{iH_{\rm S}\tau}\rho(t)e^{iH_{\rm S}\tau}.
\end{align}
By this approximation, Eq.~(\ref{eq:Red1}) can be made local in time
\begin{align}
    \frac{d}{dt}\rho^{\rm I}(t)= g^2 \int^t_0d\tau \Big( 
    -S^{\rm I}(t)C(\tau)S^{\rm I}(-\tau)\rho^{\rm I}(t)
    +S^{\rm I}(t)\rho^{\rm I}(t)C^{*}(\tau)S^{\rm I}(-\tau) 
    \Big) +{\rm h.c.}. \label{eq:Red2}
\end{align}

\section{Simple example of proof \label{sec:example1}}
\renewcommand{\theequation}{B.\arabic{equation}}
\setcounter{equation}{0}
The basic idea to prove that $\Lambda_t^{{\rm l}, {\cal B}}$ and $\Lambda_t^{{\rm nl}, {\cal B}}$ are CP can be grasped by considering a simple example of $\Lambda_t^{{\rm l, ex}, {\cal B}}$.
The first-order, linear differential equation 
\begin{align}
    \frac{d}{dt}\Lambda_t^{{\rm l, ex}, {\cal B}}\rho=L \Big( \Lambda_t^{{\rm l, ex}, {\cal B}}\rho     \Big)L^{\dagger} \label{eq:1B1}
\end{align}
has the formal solution
\begin{align}
    \Lambda_t^{{\rm l, ex}, {\cal B}}\rho= \rho + \sum_{n=1}^{\infty}\frac{t^n}{n!}L^n (\rho) \big(L^{\dagger} \big)^n.\label{eq:1B2}
\end{align}
Taking the time derivative of Eq.~(\ref{eq:1B2}) shows that it satisfies Eq.~(\ref{eq:1B1})
\begin{align}
    \frac{d}{dt}\Lambda_t^{{\rm l, ex},{\cal B}}\rho= \sum_{n=1}^{\infty}\frac{t^{(n-1)}}{(n-1)!}L^n (\rho) \big(L^{\dagger} \big)^n= L \Big( \rho + \sum_{n=1}^{\infty}\frac{t^n}{n!}L^n (\rho) \big(L^{\dagger} \big)^n    \Big)L^{\dagger}. 
\end{align}
The CP of $\Lambda_t^{{\rm l, ex},{\cal B}}$ is guaranteed by the positive semidefiniteness of ${\cal M}(\Lambda_t^{{\rm l, ex},{\cal B}})$
\begin{align}
    {\cal M}(\Lambda_t^{{\rm l, ex},{\cal B}})=|\langle \Psi|\Phi\rangle|^2+\sum_{n=1}^{\infty}\frac{t^n}{n!}|\langle \Psi|L^n|\Phi\rangle|^2\geq0.
\end{align}

\section{ $V_t$ and $V_t^{-1}$ \label{sec:Vtinverse}}
\renewcommand{\theequation}{C.\arabic{equation}}
\setcounter{equation}{0}
In this section, the inverse operators $V_{t}^{-1}, $ and $ (V_{t}^{\dagger})^{-1}$ are shown to exist for $V_{t}, $ and $V_{t}^{\dagger}$ defined in Subsec.~\ref{sec:localZ}.
Here, we introduce operators $V_{t,t_0}$ and $ V_{t,t_0}^{\dagger}$ and their alternative expressions, which become $V_t,$ and $V_{t}^{\dagger}$ at $t_0=0$
\begin{align}
    V_{t,t_0}&\equiv\overleftarrow{\cal T}e^{-\int^t_{t_0}ds\int^s_{t_0}ds'W_{s,s'}}=\overleftarrow{\cal T}e^{-\int^t_{t_0}ds'\int^t_{s'}dsW_{s,s'}}, \label{eq:V1}\\
    V_{t,t_0}^{\dagger}&\equiv\overrightarrow{\cal T}e^{-\int^t_{t_0}ds\int^s_{t_0}ds'W_{s,s'}^{\dagger}}=\overrightarrow{\cal T}e^{-\int^t_{t_0}ds'\int^t_{s'}dsW_{s,s'}^{\dagger}}. \label{eq:V2}
\end{align}
Taking the time derivative of Eq.~(\ref{eq:V1}) and Eq.~(\ref{eq:V2}) with respect to time $t$ or $t_0$ explicitly shows the differences between the equivalent expressions
\begin{align}
    \frac{d}{dt}\overleftarrow{\cal T}e^{-\int^t_{t_0}ds\int^s_{t_0}ds'W_{s,s'}}&=\int^t_{t_0}ds_2\big(-W_{t,s_2}\big)\overleftarrow{\cal T}e^{-\int^{t}_{t_0}ds\int^{s'}_{t_0}ds'W_{s,s'}} , \label{eq:V12} \\
    \frac{d}{dt_0}\overleftarrow{\cal T}e^{-\int^t_{t_0}ds'\int^t_{s'}dsW_{s,s'}}&=\int^t_{t_0}ds_{2n-1}\overleftarrow{\cal T}e^{-\int^t_{t_0}ds'\int^t_{s'}dsW_{s,s'}} \big(-W_{s_{2n-1,t_0}}\big), \label{eq:V22} \\
    \frac{d}{dt}\overrightarrow{\cal T}e^{-\int^t_{t_0}ds\int^s_{t_0}ds'W_{s,s'}^{\dagger}}&=\int^t_{t_0}ds_2\overrightarrow{\cal T}e^{-\int^{t}_{t_0}ds\int^{s'}_{t_0}ds'W_{s,s'}^{\dagger}}\big(-W_{t,s_2}^{\dagger}\big), \label{eq:V32} \\
    \frac{d}{dt_0}\overrightarrow{\cal T}e^{-\int^t_{t_0}ds'\int^t_{s'}dsW_{s,s'}^{\dagger}}&=\int^t_{t_0}ds_{2n-1}\big(-W_{s_{2n-1},t_0}^{\dagger}\big)\overrightarrow{\cal T}e^{-\int^t_{t_0}ds'\int^t_{s'}dsW_{s,s'}^{\dagger}} . \label{eq:V42}
\end{align}

Then, consider the following operators $\overrightarrow{\cal T}e^{+\int^t_{t_0}ds\int^s_{t_0}ds'W_{s,s'}}=\overrightarrow{\cal T}e^{+\int^t_{t_0}ds'\int^t_{s'}dsW_{s,s'}}$ such that 
\begin{align}
    & \frac{d}{dt}\overrightarrow{\cal T}e^{\int^t_{t_0}ds\int^s_{t_0}ds'W_{s,s'}} 
= \int^t_{t_0}ds_2\overrightarrow{\cal T}e^{\int^t_{t_0}ds\int^s_{t_0}ds'W_{s,s'}}W_{t,s_2}, \label{eq:V-11} \\
    & \frac{d}{dt_0}\overrightarrow{\cal T}e^{\int^t_{t_0}ds'\int^t_{s'}dsW_{s,s'}}= \int^t_{t_0}ds_{2n-1}W_{s_{2n-1,t_0}}\overrightarrow{\cal T}e^{\int^t_{t_0}ds'\int^t_{s'}dsW_{s,s'}}, \label{eq:V-12}
\end{align}
and $\overleftarrow{\cal T}e^{+\int^t_{t_0}ds\int^s_{t_0}ds'W_{s,s'}^{\dagger}}=\overleftarrow{\cal T}e^{+\int^t_{t_0}ds'\int^t_{s'}dsW_{s,s'}^{\dagger}}$ such that
\begin{align}
    & \frac{d}{dt}\overleftarrow{\cal T}e^{\int^t_{t_0}ds\int^s_{t_0}ds'W_{s,s'}^{\dagger}}=\int^t_{t_0}ds_2W_{t,s_2}^{\dagger}\overleftarrow{\cal T}e^{\int^{t}_{t_0}ds\int^{s'}_{t_0}ds'W_{s,s'}^{\dagger}}, \label{eq:V-13}\\
    & \frac{d}{dt_0}\overleftarrow{\cal T}e^{\int^t_{t_0}ds'\int^t_{s'}dsW_{s,s'}^{\dagger}}=\int^t_{t_0}ds_{2n-1}\overleftarrow{\cal T}e^{\int^t_{t_0}ds'\int^t_{s'}dsW_{s,s'}^{\dagger}}W_{s_{2n-1},t_0}^{\dagger}. \label{eq:V-14}
\end{align}
Actually, they are the inverses of $V_{t,t_0}$ and $(V_{t,t_0})^{\dagger}$.
This can be shown by taking the time derivatives of their products such as
\begin{align}
    \frac{d}{dt}\Bigg( \overrightarrow{\cal T}e^{+\int^t_{t_0}ds\int^s_{t_0}ds'W_{s,s'}} \overleftarrow{\cal T}e^{-\int^t_{t_0}ds\int^s_{t_0}ds'W_{s,s'}} \Bigg)=0. \label{eq:V1-1} \\
    \frac{d}{dt_0}\Bigg(\overleftarrow{\cal T}e^{-\int^t_{t_0}ds'\int^t_{s'}dsW_{s,s'}}\overrightarrow{\cal T}e^{+\int^t_{t_0}ds'\int^t_{s'}dsW_{s,s'}} \Bigg)=0. \label{eq:V1-2}
\end{align}
Therefore, the quantity in the parentheses is independent of $t$ and $t_0$, which is nothing but unity. Taking $t_0=0,$ $V_t^{-1}$ is shown to be well defined as the inverse of $V_t$.
The same discussion holds for $V_t^{\dagger}$.

\printbibliography

\end{document}